# Monopoly Power on the Web
## ⸻ A Preliminary Investigation of Search Engines

Tair-Rong Sheu, Kathleen Carley

Department of Engineering and Public Policy / Carnegie Mellon University

29[th] Telecommunications Policy Research Conference
October 27, 2001

**Abstract:**

    E-Commerce challenges traditional approaches to assessing monopolistic practices due to the rapid rate of growth, rapid change in technology, difficulty in assessing market share for information products like web sites, and high degree of interconnectivity and alliance formation among corporations. This paper has provided a fundamental framework that integrates a network and economic perspective to understand IT markets. This framework was applied to the search engine market. We study critical characteristics of the search engine market, including major players, search techniques, network structure, and market structure and try to assess whether any search engines have monopoly power given that advertising price in the search engine market is not easily attained and that no price is charged from users for most services provided by search engines.

    We focus on major search engines that provide general search services. We assume that the top 19 search engines in the June 2000 rating from Nielsen/NetRatings account for 100 % market share. We collected data on the hyperlinks connecting these search engine web sites over a five-month period from August 12[th] 2000 to Dec. 12[th] 2000. Each month's network was stored as a binary matrix. By analyzing these matrices, we found that the degree of centrality has increased on average from August to December, although the trend is not monotonic. There is also a fairly high degree of variance, regardless of the centrality measure used. This indicates that there is an unequal distribution of power in this industry.

    We also apply three additional concepts: concentration, product differentiation, and entry barriers to describe the market structure of the search engines. These are important measures used by the Department of Justice to evaluate markets. The findings indicate

that (1) despite an increasing number of search engines, barriers to entry seem high, largely due to the exponential growth in the number of web sites and the non-scalability of the current search technology and collective switching costs; (2) older search engine sites tend typically to have more features to lock in users. Using standard economic indicators (CR4=58% and HHI=1163), the industry looks close to being plagued by anticompetitive practices. However, based on a network adjusted HHI (NAHHI) constructed in this paper, its value, 870, suggests that there is less cause for concern.

To date, no search techniques are able to cover the whole web. Estimates suggest that the most comprehensive search engines are covering less than half of the sites in 2000. The number of sites is increasing faster than the number of sites searched. If this continues, and if the number of search engines decreases, then likely results are information distortion, loss of social welfare, and loss of economic value for sites not covered by the search engines. If there were to be only a few search engine sites, who are they likely to be? Based on all indicators, it suggests that Yahoo would be a contender. Other possible contenders are MSN and Netscape. On the basis of results to date, some search engines keep increasing their audience reach while others don't. The trend shows that some search engines may dominate the search engine market. We suggest conducting research in the coverage performance of search engines and investigate "information search cost" as a performance indicator of search techniques. In addition, we suggest paying attention to any anticompetitive conduct (e.g. product bundling) that may lesson competition and reduce consumer welfare. The combination of network theory and economic theory to study the search engine market, used in this paper, is a particularly powerful approach for E-Commerce.

Contact information:

Tair-Rong Sheu: tsheu@alumni.carnegiemellon.edu

Kathleen Carley: kathleen.carley@cmu.edu

**Acknowledgement: We thank Dr. Scott Farrow for his great assistance to this paper.**

**Introduction:**

According to a report by Cyveillance (2000), as of June 2000, there were 2.1 billion indexable pages[1] on the web. By Jan. 2001, 4 billion pages are expected. People find web pages via search engines (Nielsen Study, 1997; Ernst & Young, 1998; GVU, 1998; NetRatings, 1999; IMT Strategies, 2000), web sites that assist Internet users to locate other web sites (domain names) or pages (Uniform Resource Locator, URL). Smith, Bailey, and Brynjolfsson (1999) observed that the search cost for individual users increases due to the sheer volume of information. Search engines help users reduce search costs to find web pages. Therefore, search engines are the "portals" to the rest of the Web and thus attract high hit rates. Of the top 10 web sites studied by MediaMetrix (2000), nine provide general search services[2].

Establishing a web site on the Internet is very easy; however, not every site gains equal attention. Adamic and Huberman (1999) in a study of log files from AOL found that the top 5 % of sites attract 75 % of the users. Hence, a small number of sites command the traffic of a large segment of the web population. This is a characteristic of winner-take-all markets (Frank and Cook, 1995). According to a study in June 2000 from Neilsen/NetRatings[3], Yahoo had 47 % of audience reach[4], which means 47 % of the survey participants had been to the Yahoo web site during the survey period. After Yahoo, MSN, Go, and Netscape had 35 %, 19 %, and 15 % audience reach, respectively[5]. These numbers indicate high concentration as defined by audience reach. They also showed that Yahoo and MSN have continuously increased their audience reach while most other search engines have decreased. If the Web is a winner-take-all market, then some web sites will dominate the market. To date, some of the search engines have gained high hit rates (Nielsen/NetRatings, 2000) and may dominate the market. Do search engines have monopoly power?

Monopoly power is a key factor in antitrust cases. However, antitrust related issues on the Web have not gained significant attention. Rather, attention has been focused on issues such as pricing, trust, and loyalty (Bailey, 1998; Kollock, 1999; Telang, Mukhopadhyay, and Wilcox, 2000). Sheremata (1998) stated that antitrust policy and enforcement in information technology (IT) industries appear to be difficult. The dynamics of competition and industrial organization in these industries are difficult to understand, due to the special nature of the new economy and to the scarcity of research about the IT industries (Sheremata, 1998). The Microsoft antitrust case has caused many discussions about the characteristics and conduct of these industries as well as debates over antitrust policies and remedies in the IT industries. Nevertheless, the nature of competition on the Web is still not clearly understood. Since search engines attract significant attention from users, studies of search engines are popular (Nielsen Study, 1997; Ernst & Young, 1998; GVU, 1998; NetRatings, 1999; IMT Strategies, 2000), but they do not provide an understanding of the industry. This paper tries to understand the search engine market from a perspective that integrates technology, behavior, economics, social networks, and organizational theory. This enables a more comprehensive evaluation of whether or not search engines have monopoly power. The conduct and performance of search engine sites will be discussed. Since the Web is still in its early stage of development, understanding the technology and behavior of the Web can help policy makers determine the relevance of existing antitrust legislation to E-Commerce.

## 1 Motivation: Microsoft Case

### Background Information and Debates:

The Department of Justice (DOJ), 20 state attorneys general, and the District of Columbia filed broad antitrust lawsuits against the Microsoft Corporation on May 18, 1998, charging that Microsoft had illegally thwarted competition to protect and extend its monopoly over personal computing software. Essentially, the government contended that Microsoft was violating the Sherman Antitrust Act of 1890 by using its monopoly of Windows 95 to dominate the Internet browser market[6].

As Microsoft's conduct was somewhat different from standard antitrust behavior with respect to monopoly and anticompetitive activities, it became a controversial question whether Microsoft was involved with antitrust issues. The pros and cons of antitrust action against Microsoft (Spaulding, 2000; Thierer, 1997; the Chicago Tribune, 1997) in terms of the key issues of monopoly power, price, quality, innovation, and competition are listed in Table 1.

Table 1: The Pros and Cons of Debates in Microsoft case

| | Pros (Not Antitrust) | Cons (Yes-Antitrust) |
|---|---|---|
| Monopoly Power | In 1997, Microsoft only accounted for less than 2% of the entire computer hardware and software industry, 4% of the software industry, and 36% of the browser market. | Microsoft dominates 90% of the Operating Systems (OS) market. This gives it a natural monopoly due to network effects in the software industry. The marginal cost of software is near zero. Microsoft has established a huge installed base that gives Microsoft power to price. |
| Price | Predatory pricing of software is impossible because competition drives costs to slightly above marginal cost, and since the marginal cost of software is nil, it's impossible to price below cost. According to this, Microsoft was just selling Internet Explorer (IE) at marginal cost. Also the theme that Microsoft would raise the price of IE after it drove all its competitors is based on traditional antitrust theory and will not happen because new competitors will likely enter the market. | Microsoft used predatory pricing to establish its monopoly power in the browser market. If the marginal cost of IE is nil, the marginal cost of all software should be nil. The reason why IE is free and other software (including operating systems) is so expensive is that IE faces real competition while Microsoft has already monopolized markets in other areas. |
| Quality | Microsoft keeps improving its products. The functions of the latest version of IE are much better than its initial versions. Microsoft developed the most desirable product and continues to build on its success with innovation and imagination. | Microsoft rarely produces best software in the beginning. Their software might be considered good because users don't see any other products. IE might be better than Netscape now. That is because Microsoft spends money made from its other monopoly markets on developing IE and gives it away for free. Netscape can't make money after Microsoft's competition resulting in poor R&D development. |
| Innovation | Microsoft keeps innovating. The tying of IE with the OS is evidence that Microsoft can offer better functions by integrating the two products together. | Once Microsoft monopolizes the browser market, it will be able to set the standard and develop proprietary technologies, hindering innovation by other companies. |
| Competition | The AOL-Netscape merger will give Microsoft more competition. | Not really. Microsoft isn't charging for IE separately. AOL-Netscape won't be able to affect Microsoft's monopoly positions in its markets for OS, Office, and so on. |

The Microsoft case illustrates that in IT industries the case for and against monopolistic practices is somewhat different than in non-IT industries. Microsoft had a monopoly (90% of market share in the Operating System market), high production differentiation (hard to find similar products), and there was a high entry barrier. In addition, Microsoft was involved with many computer manufacturers and Internet service/content providers and had forced them to conduct anticompetitive acts.

Microsoft's network position in the IT industry is in fact similar to that of other organizations that have been involved in antitrust suits. Chowdhury (1998) examined change in the inter-organizational network within the IT industries from 1989 to 1997. He found that the 4 companies with the highest degree of centrality (largest number of connections[7] to other companies) were AT&T, IBM, Microsoft, and Time Warner. All four have been involved with antitrust issues. Chowdhury found that after 1993 Microsoft dramatically increased the number of other organizations with which it was allied (i.e., the degree of centrality increased). This was correlated with Microsoft's growth on other fronts and its increase in market share. Indicators based on analyses of the inter-organizational network structure and the market economy support the claim that Microsoft had monopolistic power. The search engine industry is another IT industry. Will we see a similar pattern of behavior to that observed in the Microsoft case?

## 2 Research Model

A standard approach to analyze markets used by Industrial Organization Economists (Viscusi, Vernon, and Harrington, 1995) is the Structure-Conduct-Performance (SCP) model (Figure 1 – non-shaded portion). This model provides a framework for organizing and discussing the important concepts of markets. Implicitly, this model assumes that connections among organizations are irrelevant, and that organizations are independent actors.

A complementary approach has been taken by organizational theorists, who argue that organizations are not independent; rather, the market is a social network (e.g. White, 1981a, 1981b; Baker, 1984a, 1984b; Faulkner, 1983). Further, this inter-organizational network affects the performance of organizations and their ability to constrain and enable the deployment of new products and the entry of new organizations (Burt, 1982; Leifer and White, 1987).

Figure 1: Standard SCP model (unshaded) and Extended Model (shaded)

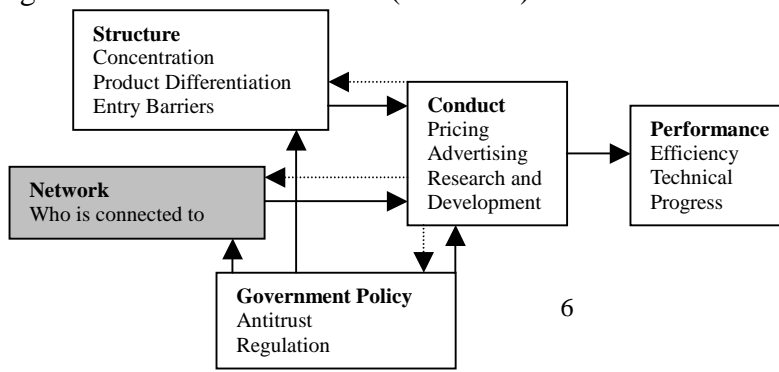

Baker (1987) further argued that when this inter-organizational network was differentiated into a core and periphery structure, there was an unequal distribution of power. Such core-periphery structures are another potential indicator of monopolistic control or collusion. Core actors would have a much higher degree of centrality than non-core actors.

If core actors control important resources in the economy, they possess the means to control economic decisions (Parsions and Smelser, 1956). Core actors tend to be more profitable (Burt, 1983). Social network relationships not only help firms control resources in order to affect economic decisions, but also help firms overcome the uncertainty and distrust that often plague market exchange (Granovetter, 1985). Interconnection among firms facilitates the flow of resources and information, but also increases the possibility that firms who are connected as a subnetwork pursue their interests, such as increasing price, reducing competition, deterring new entrants, etc. Taking these inter-organizational networks into account leads to an extension of the SCP model as shown in Figure 1 with the addition of the shaded box.

## 3   Preliminary Investigation of Search Engines

### 3.1 Definition of the Market of Major Search Engines

Although there are several types of search engines[8], in this paper, we focus on the major search engines that provide general search services. We assume that the top 19 search engines[9]  in the June 2000 rating from Nielsen/NetRatings account for 100 % market share. The reasons are (1) the top 19 search engines listed by Nielsen/NetRatings have gained significant audience reach (at least 0.1 %) and are used by most people; (2) the engines outside the top 19 either have insignificant market share (less than 0.1 % of audience reach) or are not considered a search engine by Nielsen/NetRatings. This selection of 19 organizations that come from Nielsen/NetRatings to do a market analysis may overstate the case for monopolistic effects as it does not include AOL[10], which is an Internet Service Provider website that also provides search service.

## 3.2 Characteristics of Search Engine Market

### 3.2.1 Search Techniques

Due to the fast growth of the Web and limitations of indexing and crawling techniques, none of the available search techniques can thoroughly cover the whole Web. Each technique has its own advantage and disadvantage[11]. Generally speaking, search techniques fall into five categories:

**Human Editing**

Yahoo is the best-known site in this category. Yahoo has more than 150 editors to screen and categorize URLs submitted by their owners. Yahoo emphasizes to their users that their editors review every URL in their directory. Another kind of human editing is question-based. AskJeeves, for example, also hires editors to edit web pages as answers to user questions.

**Crawler**

AltaVista and Google are examples of crawler search engines. Crawler search engines use software to create a database[12]. Although technical concepts behind these crawler search engines are similar, they can use very different search methodologies and algorithms to decide relevancy and ranking of search results. So far none of the crawler search engines can cover the whole Web. Even Google, which claimed to have a biggest indexed database, only covered half of the web as of June 2000. The reasons are (1) the scope of the Web is growing faster than the batch process in which search techniques can crawl; (2) some domains are proprietary and reluctant to be searched; (3) some search engines only search web pages in specific languages.

**Popularity**

DirectHit is widely known for its popularity methodology. DirectHit collects data about what people click on when search results are provided to them from the DirectHit and HotBot sites. The database is updated periodically. The more people click on a site, the higher ranking that web site is given.

**Commercial**

Goto is one of the "Commercial" search engines. How sites are ranked as a result of a search is based on how much they pay for each click. The basic concept is similar to the yellow pages. Commercial search engines believe that the more web sites pay for each click, the better the quality of the web sites. Therefore, when users click on higher ranked web sites, they presumably visit higher quality web sites.

**Hybrid**

Some search engines employ hybrid search techniques[13] in an attempt to increase coverage. Some search engines provide second opinions from other search engines following their initial results.

### 3.2.2 Network Structure of Search Engines

So far, none of the search techniques employed by the search organizations can adequately cover the whole web. Thus, interconnection of search engines sites is one way to increase coverage. Many search engines provide hyperlinks pointing to other search engines (Figure 2), though others, such as MSN and iWon, don't. When users click on those hyperlinks, users are automatically taken to those search engines to get the results of their search without inputting keywords again. These hyperlinks create a network of search engines. The power that sites have over the flow of information is a function of the site's position in this network. Network centrality, both degree and betweenness have been used as indicators of power.

Figure 2: Hyperlinks on Yahoo's search result pages

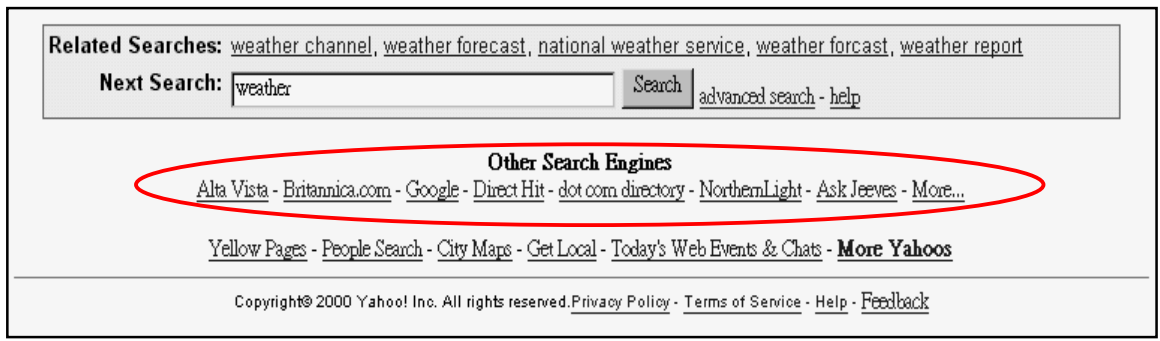

 Source: Yahoo's web page of search results, Nov. 19, 2000

Network analysis is a methodology that uses nodes and lines to map relationships of actors. Wasserman and Faust (1994) stated that the network perspective has proved fruitful in a wide range of social and behavioral science disciplines. A network is composed of nodes and relations. For our purposes, the nodes are search engine web sites. Relations are the hyperlinks, the channels for transfer or " flow" of resources (either material or nonmaterial). Nodes vary in their centrality. We use three measures of centrality – degree, betweenness, and information. The most commonly used measure of centrality is degree, the number of other nodes the node in question is connected to. Degree can be separated into indegree (the number of other search engines that connect to the engine in question) and outdegree (the number of other search engines to which the engine in question connects to).

Betweenness and information are also used to measure node centrality. Freeman (1979) uses betweenness centrality[14] to measure the extent to which a particular node lies "between" the various other nodes in the graph: a point of relatively low degree may play an important "intermediary" role and so be very central to the network (Scott, 1991). Information centrality[15] measures the information contained in all paths originating with a specific node. The information of a node averages the information in these paths, which, in turn is inversely related to the variance in the transmission of a signal from a node to another (Wasserman and Faust, 1994).

We collected data on the hyperlinks connecting search engine web sites over a five-month period from August 12[th] 2000 to Dec. 12[th] 2000. Each month's network was stored as a binary matrix. For example, when Yahoo places a hyperlink leading to AltaVista[16], then 1 is recorded in cell defined by the Yahoo row and AltaVista column. Otherwise, 0 is recorded. The data collected from August 12[th] to Dec. 12[th] were recorded in five matrixes.

We checked the 19 search engines listed in the rating from Nielsen/NetRatings in June, 2000. At each site, we used four types of keywords, including "travel" (popular word) "Citibank" (company name) "Tair-Rong Sheu" (unusual name), and "fdkhgugn" (meaningless word). Then we recorded the hyperlinks pointing to the other search engines that are in the top 19 list.

The network structure[17] of the search engines on August 12[th], 2000 is shown graphically in Figure 3. The direction of the arrow indicates who links to whom.

Figure 3: Network Structure of Search Engines on August 12[th], 2000

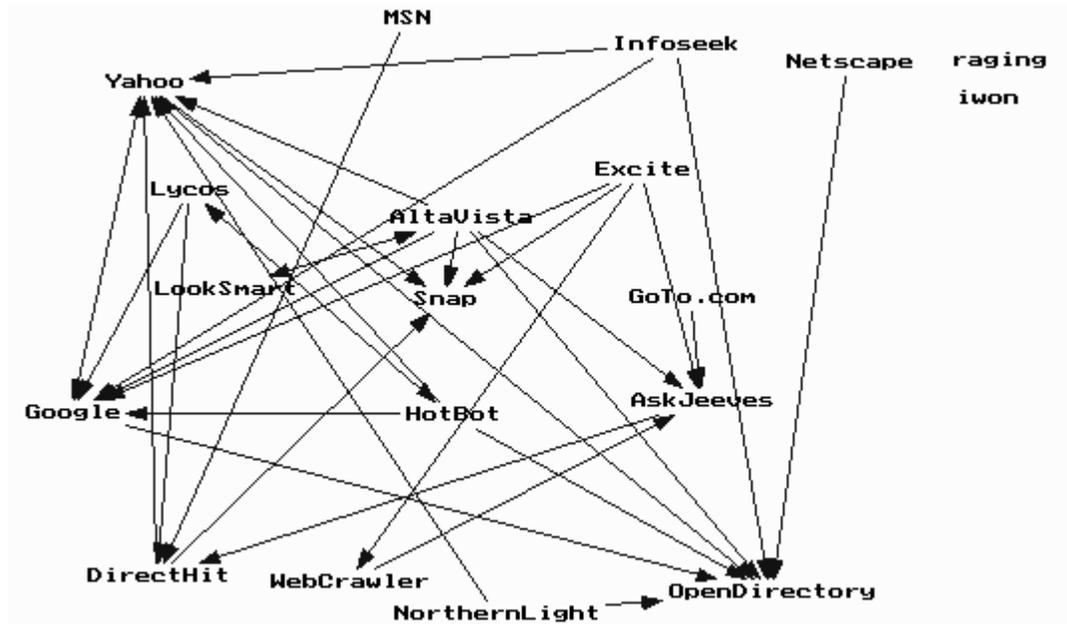

Table 2 summarizes the way in which average centrality changes over time[18]. The degree of centrality has increased on average from August to December, although the trend is not monotonic. This increase is a sign of network structuration – the maturation and development of a stable network of relations that control an industry. There is also a fairly high degree of variance, regardless of the centrality measure used. This indicates that there is an unequal distribution of power in this industry.

Table 2: Change in Average Centrality in the Search Engine Network

| Date of data collection | Aug. 12th | Sep. 12th | Oct. 12th | Nov. 12th | Dec. 12th |
|---|---|---|---|---|---|
| Mean of Indegree (Stdev*) | 1.84(1.60) | 2.26(1.89) | 2.05(1.73) | 2.11(1.74) | 2.32(1.72) |
| Mean of Outdegree (Stdev) | 1.84(2.32) | 2.26(2.69) | 2.05(2.63) | 2.11(2.67) | 2.32(2.75) |
| Mean of nBetweenness (Stdev) | 0.96(1.80) | 1.08(1.75) | 0.83(1.49) | 0.76(1.37) | 1.50(2.41) |
| Mean of Information Centrality (Stdev) | 0.44(0.14) | 0.51(0.16) | 0.48(0.15) | 0.48(0.15) | 0.49(0.16) |
| | | | | | |
| Density | 0.102(0.303) | 0.126(0.332) | 0.114(0.318) | 0.117(0.321) | 0.129(0.335) |
| Sample size | 19 | 19 | 19 | 19 | 19 |

*: Stdev: Standard Deviation. Others apply.

We compare the node behaviors of the top 4 search engines (Yahoo, MSN, Go, Netscape, decided by their audience reach) to all other engines (see Figure 4). The top 4 sites have lower indegree and higher outdegree than do the other sites. In other words, these 4 sites dominate in terms of directing users where to look for information, both directly and indirectly. During the five-month period, Yahoo and Netscape have an average outdegree of 5.2 and 6.4, respectively, but MSN and Go (Infoseek) have none. This reflects two different business strategies. Yahoo and Netscape have the most information power.

Figure 4: The Indegree and Outdegree Trend of Top 4 v.s. All Others (15)

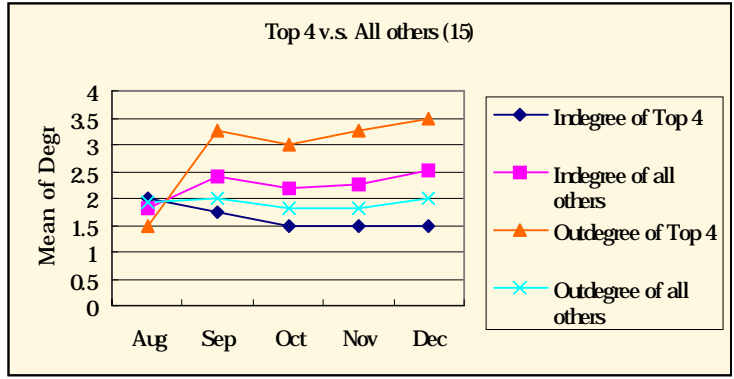

The top 4 sites with the highest centrality for each network measure are shown in Table 3. As we saw in the case of Microsoft, both a highly central network position and market share (audience reach) are needed to exert monopolistic influence. The network-based approach suggests a somewhat different view of power than does the audience reach measure. Yahoo is among the top 4 sites on all measures. Thus, from both a network position and an audience reach position, Yahoo has the potential to have the most monopolistic influence. MSN and Go, although high in audience reach, have no centrality. If resource based competition enables monopolistic power, then there may also be antitrust issues surrounding MSN. Table 3 also shows that some search engines (e.g. Alta Vista, Google) have highly central network positions but their audience reach are not among top 4. The longitudinal data of audience reach ranking after Aug. 2000 are needed to verify whether a central network position leads to high audience reach in the future.

Table 3: Top 4 Sites Under Each Metric of Power Using August 2000 Data

| Indegree | Outdegree | Betweenness (normailzed) | Information (normailzed) | Audience Reach in June 2000 |
|---|---|---|---|---|
| AltaVista(6) | Open Directory(7) | DirectHit(5.66) | Yahoo(0.56) | Yahoo(47%) |
| Excite(4) | Google(6) | Yahoo(4.90) | Google(0.56) | MSN(35.8%) |
| HotBot(4) | Yahoo(6) | AskJeeves(4.25) | Alta Vista(0.55) | Go(19.1%) |
| Go,Lycos, Yahoo(3) | Ask Jeeves, Snap(4) | Alta Vista (1.96) | Open Directory(0.55) | Netscape(15.4%) |

### 3.2.3 Standard Market Structure of Search Engines

Three additional concepts: concentration, product differentiation, and entry barriers, are used to describe the market structure of the search engines. These are important measures used by the Department of Justice to evaluate markets. We will later compare the characteristics of the search engine market to the criteria listed in 1992 Horizontal Merger Guidelines[19] and discuss policy implications.

The market structure analysis is conducted using the top 19 search engines[20] listed by Nielsen/NetRatings. We make two assumptions: (1) the top 19 search engines account for 100 % of market shares[21]; (2) the market sales[22] are linearly proportional to audience reach.

Nielsen//NetRatings has kept tracking the audience reach of top search engines for more than one year (Figure 5). There were 19 search engines on Nielsen//NetRatings as of June 2000, from the number one ranked Yahoo with 47 % of audience reach to the number nineteen Raging with 0.1 % of audience reach[23].

Figure 5: The Trend Comparison of Audience Reach of Search Engines

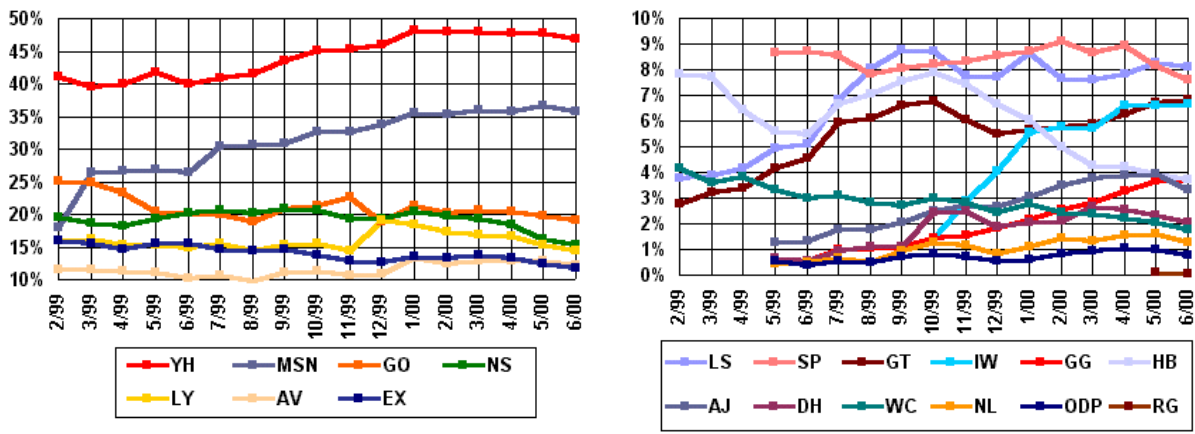

Source: www.searchenginewatch.com

KEY: YH=Yahoo, MSN=MSN, GO=Go (Infoseek), NS=Netscape, LY=Lycos, AV=AltaVista, EX=Excite,    LS=LookSmart,
SP=Snap, GT=GoTo, IW=iWon, GG=Google, HB=HotBot, AJ=AskJeeves, DH=Direct Hit, WC=WebCrawler, NL=Northern Light,
ODP=Open Directory, RG=Raging Search

### Concentration

To determine concentration, three measures are used: the four-firm concentration ratio, the Herfindahl-Hirschman Index, and the Network Adjusted Herfindahl-Hirschman

Index.

- Four-firms concentration ratio (CR4) from Jul. 1999 to Jun., 2000

A concentration ratio is the most widely used index to measure concentration (Viscusi, Vernon, and Harrington, 1995). The four-firms concentration ratio reflects the share of total industry sales accounted for by the 4 largest firms. In order to calculate the total market share, the "market" has to be defined. It is necessary in practice to make difficult judgements about what products and firms constitute the market (Viscusi, Vernon, and Harrington, 1995).

Since we don't have actual data of the advertising market sales, we define the search engine market share as follows:

$$\text{Search engine i's market share (Pi)} = \frac{\text{Search engine i's audience reach}}{\text{Summation of total audience reach}}$$

We calculate CR4 by taking the summation of the market shares of the top 4 search engines. The CR4 in each month is about 58% from Jul. 1999 to Jun. 2000[24].

- Herfindahl-Hirschman Index (HHI)

HHI is an index used by the Department of Justice and the Federal Trade Commission in their 1992 Horizontal Merger Guidelines. The HHI has the advantage of incorporating more information showing the size distributions of firms than the concentration ratio does. HHI is defined as:

$$\mathbf{HHI} = \sum_{i=1}^{N} (100 Pi)^2$$

Pi is the firm i's percentage of total industry sales (i.e., its market share), where N is the number of firms. The HHI ranges from 10,000 (a pure monopoly market) to a number close to zero (an atomistic market).

There are 19 search engines included in this calculation (N=19; Pi is defined above). The average HHI is 1163 over the period from Jul. 1999 to Jun. 2000[25].

- Network Adjusted Herfindahl-Hirschman Index (NAHHI)

We propose to use NAHHI to evaluate the concentration of web sites. Hyperlinks placed on web pages form a network and facilitate users visiting from one site to another. Therefore, the market share (audience reach) of a site should incorporate the audience that may follow hyperlinks to visit it. If search engines are not connected to each other, the NAHHI is equal to HHI. If search engines are connected to each other, users may follow those links to visit other search engines and become aware of and familiar with those search engines. NAHHI reflects potential market shares that search engines would acquire through interconnection and substitute information provided to users through hyperlinks. The higher the network density (interconnection), the lower the NAHHI. To calculate NAHHI, based on the indegree of search engines, we derive the "possible" audience reach (market share), which is the sum of its audience reach and the audience reach of those search engines that place a hyperlink pointing to it. Because users may visit more than one search engine (i.e., the total audience reach exceeds 100 %), we normalize the possible audience reach by taking the possible audience divided by the sum of the possible audience reach of the 19 search engines [26]. The NAHHI for August 2000 is 870, which is smaller than the HHI. The reason is that, after interconnecting to each other, the audience reach of search engines is more equally distributed. When the network effect is taken into account, there is less evidence for concentration.

**Product Differentiation**

Users often start from search engines to explore other web pages. However, the frequency with which commercial web sites are visited is proportional to revenue. One strategy that search engines use to increase visits is to offer additional services such as free e-mail, news, chat, weather, etc. This differentiates their products.

In addition to search service, products that search engines offer often fall into three product categories: (1) Non-personalized features, such as News, Weather, Stock Quote, Map, etc. For these features, personal information is not required; (2) Personalized features, such as E-mail, Online Chat, Online Game, etc. With personalized features, sites can attract more users and lock them in (Telang, Mukhopadhyay, and Wilcox, 2000); (3) Platforms, such as online shopping, auctions, etc. Older search engines (Table 4) tend to offer products other than just search service. This increases users' switching costs. The top 4 search engines tend to develop personal features (Table 5).

Table 4: Product Differentiation of Old and New Search Engines Using August 2000
Data

|  |  | Setup Dates | Number of SE | Non-personalized Features | Personalized Features | Platforms |
|---|---|---|---|---|---|---|
| Top 19 Search Engines | Old | 1994-1997 % | 14 | 13 93% | 12 86% | 13 93% |
|  | New | 1998-2000 % | 5 | 3 60% | 4 80% | 4 80% |

Table 5: Product Differentiation of Top 4 and All Others

|  | Number of SE | Non-personalized Features | Personalized Features | Platforms |
|---|---|---|---|---|
| Top 4 search engines | 4 | 4 100% | 4 100% | 4 100% |
| Top 5-19 Search Engines | 15 | 12 80% | 12 80% | 13 87% |

Logistic regressions of the presence or absence of three different product features
(1 = present, 0 = absent) onto the year of startup (1994-2000) show that the earlier the
setup date of a search engine, the higher the probability of providing products other than
search service that attract and lock in users (Table 6).

Table 6: Descriptive Statistics of Logistic Regressing Product Features onto Years of
Setup

|  | Non-personalized Features | Personalized Features | Platforms |
|---|---|---|---|
| Chi-square p-value | 0.03 | 0.04 | 0.06 |
| Odds ratio | 0.35 | 0.43 | 0.50 |
| Maximum rescaled R2 | 0.48 | 0.38 | 0.30 |

The predicted probability curves of three different product features are shown in
Figure 6-1 to 6-3.

Feature 6-1: The Non-Personalized Features of Top 19 Search Engines by Setup Dates

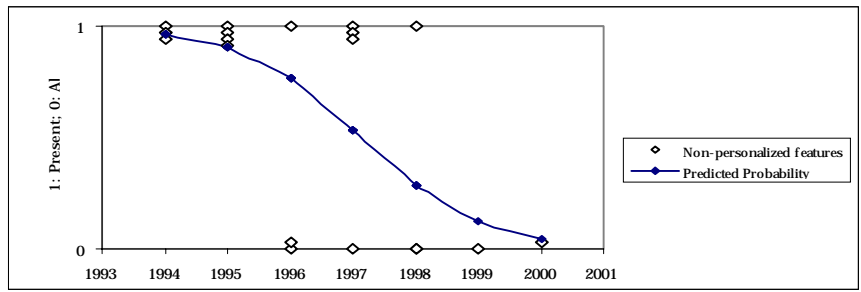

Feature 6-2: The Personalized Features of Top 19 Search Engines by Setup Dates

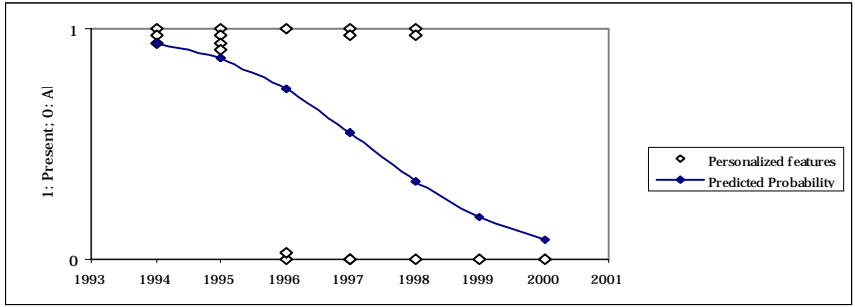

Feature 6-3: The Platforms of Top 19 Search Engines by Setup Dates

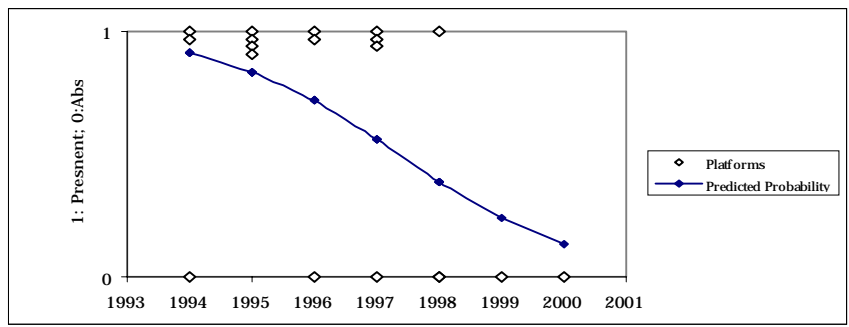

**Entry Barriers**

The entry barrier of running a search engine is increasing over time due to the increasing size of the web, and increased switching costs resulted from products (e.g. e-mail) that lock in users. In the previous section, we observed that many search engines provide products that function to lock in users. Since we haven't empirically examined the locking effect in this paper, in the following we provide a qualitative argument over entry barriers.

● **Web Size**

From Dec.1999 to June 2000, the number of web pages has increased from 1 billion to 2 billion. By Jan. 2001, 4 billion web pages are expected. Each web page is estimated to have an average of 5.6 external links (Cyveillance report, 2000). Search techniques that rely on analysis of the pages (such as human editing) and those that rely on crawling among the links, have an increasingly hard job. In 2000, Yahoo had about 150 editors and Open Directory had more than 30,000 volunteers to edit their directories. As the number of sites increases, the number of human editors needed increases. For a new engine, all the current sites would need to be instantly indexed; whereas, existing engines only need to make incremental changes. The dramatic growth of the web creates a barrier to entry. Buying a database from another search engine is one way of overcoming this barrier. However, rivalry search engine companies may jealously guard their databases.

- **Switching Costs**

The revenues of search engines are mainly from two sources: advertising and electronic commerce. Both are positively proportional to the number of users. For new entrants, developing a large base of users takes time, a good product, and a lot of promotion expenses. The difficulty arises from high collective switching cost[27] (Shapiro and Varian, 1998). Telang, Mukhopadhyay, and Wilcox (2000) showed that if users have used a particular search engine frequently in the past, they are much more likely to choose that search engine again in the future. Therefore, merging with an existing site with many users is a faster and cheaper alternative. Two merger cases showed that the average cost of acquiring an e-mail user[28] in 1998 was $40 and the average cost of acquiring a hosted web site[29] was $1000 in 1999. These investments are independent of search techniques. As previously noted, major search engines provide many features other than just search service. A new entrant has to overcome the challenge of providing competitive search service and has to add many features to attract users who have been locked in by current sites – which takes time. The longitudinal data (Figure 5) show that the top 2 sites, Yahoo and MSN, have continuously increased their audience reach while most other search engines have decreased. Although some new engines (e.g. iWon) entered the market, the CR4 over a year (note 24) is still not affected.

Figure 7 shows the distribution of the top 19 search engines by their setup dates. Of the top 19 search engines, 17 have been set up for two or more years. A simple regression on audience reach and years of setup shows that the earlier the setup date of a search engine, the higher audience reach (p = 0.05; β = 0.03).

Figure 7: The Audience Reach of Top 19 Search Engines by Setup Dates

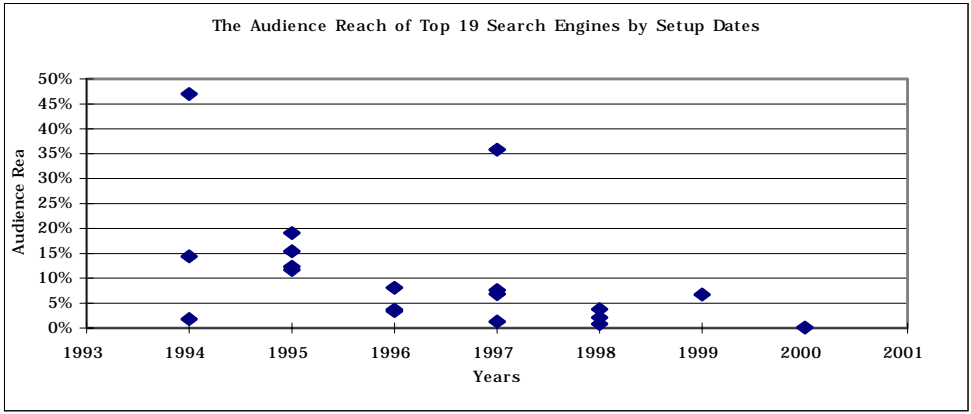

## 4    Policy Implications & Future Directions

In this paper, we have tried to assess whether search engines have monopoly power by investigating the network structure and the market structure. In order to understand whether the structure of the search engine market is potentially problematic, we compare the characteristics of the search engine market to 1992 Horizontal Merger Guidelines (Table 7). These guidelines describe some analytical foundations of merger enforcement and provide guidance enabling the business community to avoid antitrust problems when planning mergers.

Table 7: Comparison of Search Engine Market and 1992 Merger Guidelines

| 1992 Merger Guidelines | Search Engine Market |
|---|---|
| Market Definition | 19 major search engines that provide general search. |
| Market Shares | |
| Sales or capacity of firms in the defined market | 1.  Audience Reach is assumed to be market shares; <br> 2.  The audience reach of top 19 search engines is assumed to account for 100 % market share. |
| Market Concentration | |
| Herfindahl-Hirschman Index <br> 1. Unconcentrated: HHI below 1000; <br> 1.  Moderately concentrated: HHI between 1000 and 1800 <br> 3. Highly concentrated: HHI above 1800. | HHI: 1163 (in average from July 1999 to June 2000) <br> CR4: 58% (in average from July 1999 to June 2000) <br> NAHHI: 870 (in Aug. 2000) |
| Entry | |
| 1.  Timely: Whether entry can achieve significant market impact within a timely period <br> 2.  Likely: Whether entry can be a profitable, hence, a likely response to a merger having competitive effects of concern. <br> 3.  Sufficient: Whether timely and likely entry would be sufficient to return market prices to their premerger levels. | 1. Among top 19 search engines, 17 search engines are older than 2 years. <br> 2. iWon entered into this market in Oct. 1999 and achieved 6.7% of audience reach in June, 2000, ranked as 11[th]. <br> 3. Search engines don't charge users. However, price for advertising may be an indicator to measure competition but not studied in this paper. |

According to the guidelines, the search engine market is moderately concentrated based on the HHI. If there were any merger that would cause HHI to increase more than 100 points, then the merger would raise significant competitive concerns[30] and need further investigation. In Figure 8, the gray cells indicate that there are 29 possible mergers that could result in HHI increasing by more than 100 points.

Figure 8: Possible Mergers Producing an Increase in the HHI* of More Than 100 points

| Rating | 1 | 2 | 3 | 4 | 5 | 6 | 7 | 8 | 9 | 10 | 11 |
|---|---|---|---|---|---|---|---|---|---|---|---|
| SE | YH | MSN | GO | NS | LY | AV | EX | LS | SP | GT | IW |
| 1 YH | | | | | | | | | | | |
| 2 MSN | 1131 | | | | | | | | | | |
| 3 GO | 519 | 418 | | | | | | | | | |
| 4 NS | 395 | 315 | 192 | | | | | | | | |
| 5 LY | 395 | 315 | 192 | 163 | | | | | | | |
| 6 AV | 307 | 244 | 145 | 122 | 122 | | | | | | |
| 7 EX | 336 | 267 | 160 | 135 | 135 | 116 | | | | | |
| 8 LS | 195 | 153 | | | | | | | | | |
| 9 SP | 192 | 150 | | | | | | | | | |
| 10 GT | 166 | 130 | | | | | | | | | |
| 11 IW | 163 | 127 | | | | | | | | | |

*: The HHI calculation is based on the audience reach data in Jun. 2000.

However, the market would be considered unconcentrated if NAHHI were taken into account. It shows that interconnection of search engines reduces the possibility of getting a concentrated market because audience reach may be more equally distributed through interconnection.    Since the search engine industry is concerned with the flow of information among sites, the NAHHI may provide a more accurate reflection of actual concentration.

The statistical data of the top 19 search engines show that in most cases, the date of setup is critical to success. However, some new engines (e.g. iWon), with special promotion, can still achieve high audience reach in a short time. Therefore, new entrants may still affect markets and prevent anticompetition. Whether this will continue to be the case as the exponential growth of the web continues remains to be seen.

Several conducts of search engines are relevant to the investigation of anticompetition:

● **Search engines don't charge users but advertisers.**

Search engines provide free search services to users but make money by selling web spaces to advertisers. Therefore, advertising sales and prices are relevant to evaluate monopoly power of search engines so as to assess social welfare. Unfortunately at this point, data on advertising sales are difficult to collect from public sources because not every search engine is listed on the stock market.

- **Some search engines don't connect to other search engines.**

According to the study of network structure, we found that 17 of the 19 search engines connect to other search engines. We also found that the interconnection of search engines has an upward trend. However, some search engines such as MSN and iWon don't connect to any other search engines.

- **Some search engines are supported by giant parent companies and may be involved with product bundling[31].**

It is well known that Microsoft bundled the IE browser with its OS and therefore was alleged to be anticompetitive. Based on my study, we find similar situations (e.g., Microsoft bundles its IE browser with its MSN web site) If, with the help from other monopoly markets, those web sites gain monopoly powers in the search engine market, they would be able to charge advertisers higher price and might even be able to charge users search services. They might also reduce the diversity of product offered once they no longer face strong competition. Such activities could cause a loss of social welfare.

Besides conduct, performance evaluation is also a critical factor in the SCP model. Quality of search techniques may be used to evaluate performance of search engines whose core service is search, in terms of ranking and relevancy of search results. The quality of ranking and relevancy is inversely proportional to "information search cost", the cost of looking for information. Therefore, further research could investigate "information search cost" as a performance indicator of search techniques. This indicator becomes very important when the amount of information on the Web skyrockets to a degree that makes it difficult decide ranking and relevancy of search results. The situation may be even worse when ranking and relevancy of search results are not decided based on impartial algorithms[32]. As this began to happen, could users know and would they switch to other search engines? Telang, Mukhopadhyay, and Wilcox (2000) showed that users are likely to continue to use the same search engine that they have frequently used in the past. They also show that users have more loyalty to a search engine in which they use more personalized features. If search engines keep working on locking in users without improving search techniques or algorithms, this could pose a serious problem. Only a few search engines might exist, providing lower quality and less impartial service than is technically feasible.

## 5 Conclusion

E-Commerce challenges traditional approaches to assessing monopolistic practices due to the rapid rate of growth, rapid change in technology, difficulty in assessing market share for information products like web sites, and high degree of interconnectivity and alliance formation among corporations. This paper has provided a fundamental framework that integrates an economic and network perspective to understand IT markets. Additional work needs to be done to integrate these perspectives and to exam other types of alliance structures and resources. In the future, research is needed on the information search cost of search engines and on technological progress in the information technology industries.

This framework was applied to the search engine market. We study critical characteristics of the search engine market, including major players, search techniques, network structure, and market structure and try to assess whether any search engines have monopoly power given that advertising price in the search engine market is not easily attained and that no price is charged from users for most services provided by search engines. If any anticompetition is alleged in the search engine market, the research findings of this paper will be relevant. The findings indicate that (1) despite an increasing number of search engines, barriers to entry seem high, largely due to the exponential growth in the number of web sites and the non-scalability of the current search technology, and collective switching costs; (2) older search engine sites tend typically to have more features and so greater ability to lock in users. In fact, most users rarely use more than two sites in a single search session. Using standard economic indicators (CR4=58% and HHI=1163), the industry looks close to being plagued by anticompetitive practices. However, the nature of the industry and the relevant technology is such that the web of connection and alliances among search engines is highly critical both to the product being delivered and the way in which business is conducted. Thus, we constructed a network adjusted HHI (NAHHI) and its value was only 870, suggesting that there is less cause for concern.

To date, no search techniques are able to cover the whole web. Estimates suggest that the most comprehensive search engines are covering less than half of the sites in 2000. The number of sites is increasing faster than the number of sites searched. If this continues, and if the number of search engines decreases, then likely results are information distortion, loss of social welfare, and loss of economic value for sites not covered by the search engines. If there were to be only a few search engine sites, who are they likely to be? This analysis suggests that Yahoo would be a contender. It ranks high on all indicators and it is an old site, offering a variety of services, with a powerful network position and high audience reach. Other possible contenders are MSN and Netscape. On the basis of results to date, some search engines keep increasing their audience reach while others don't. The trend shows that some search engines may dominate the search engine market. We suggest conducting research in the coverage performance of search engines and investigate "information search cost" as a performance indicator of search techniques. In addition, we suggest paying attention to any anticompetitive conduct (e.g. product bundling) that may lesson competition and reduce consumer welfare. The combination of network theory and economic theory to study the search engine market used in this paper is a particularly powerful approach for E-Commerce.

**Notes**

[1] The following figure was in Cyveillance Press, July 10, 2000
http://www.cyveillance.com/newsroom/pressr/000710.asp

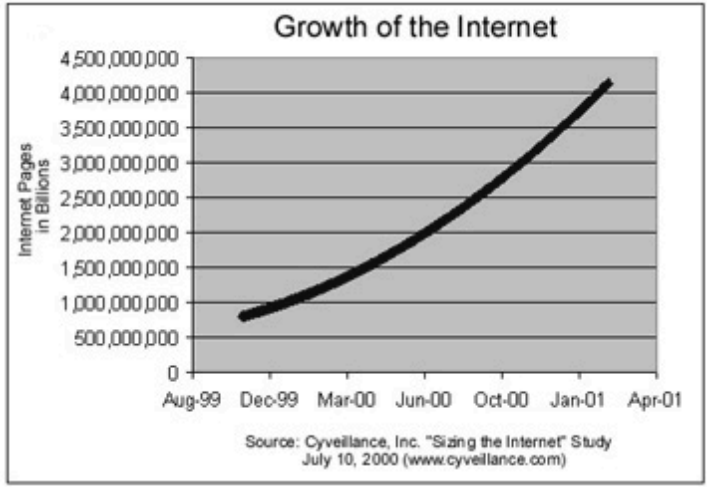

[2] The top10 web groups from MediaMetrix survey as of June, 2000 were (1) AOL Network; (2) Microsoft Sites; (3) Yahoo Sites; (4) Lycos; (5) Excite; (6) Go Network; (7) About.com Sites; (8) AltaVista Network; (9) NBC Internet; (10) Amazon. Except Amazon, the other web sites all provide general search services. The data source is from http://us.mediametrix.com/press/releases/20000720a.jsp.

[3] Nielsen//NetRatings uses software to monitor a panel of web surfers, a sample of about 43,000 at home users. The estimate of audience reach is the percentage of active web surfers estimated to have visited each search engine during the month. The data source is from www.searchenginewatch.com.
The top19 search engines surveyed by Nielsen/NetRatings in June, 2000 are shown as follows:

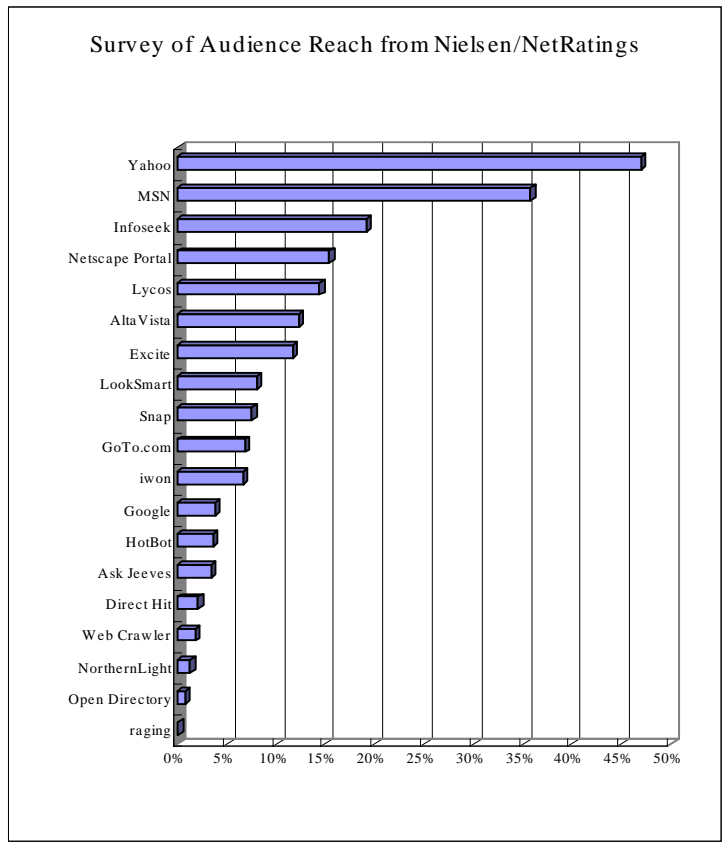

Survey of Audience Reach from Nielsen/NetRatings

[4] Audience reach, a measure used in Nielsen//NetRatings, is presented as the percentage of active web surfers estimated to have visited search engines in the survey month. One person may visit a web site several times in a given period. Thus, the number of audience reach is typically smaller than the number of hits.

[5] The combined total percentages exceed 100 % because a web surfer may visit more than one search engine.

[6] Microsoft Internet Explorer and Netscape Navigator are two major browsers in this market.

[7] The alliances in Chowdhury's research (1998) were defined as (1) merger; (2) acquisition; (3) asset purchase; (4) joint venture, joint ownership; (5) minority stake; (6) marketing or distribution alliance; (7) licensing deal; (8) supply agreement; (9) R&D partnership; (1) asset swapping; (11) partnership; (12) unclassified agreement or partnership.

[8] Some search engines that provide general search service may not be selected in my data set. The reason is that they have too little audience reach to be found by casual users and thus to have market power. We either don't include Metacrawlers, such as search.com or mamma.com, because they don't have their own directory or database. They only send

keywords to other search engines and get results back to users.

[9] Please see note5.

[10] AOL is not listed in the top 19 search engines by Nielsen/NetRatings. Whether AOL is considered a search engine is controversial since providing Internet access is its core business. However, its subscribers may perform search through its search function. Therefore, the assumption of that the top19 search engines account for 100 % of market shares may be distorted when AOL is considered a search engine and the cross-elasticity demand between AOL and other search engines is high.

[11] The advantages and disadvantages of different search techniques are shown as follows:

| Search Techniques | Advantage | Disadvantage |
| --- | --- | --- |
| Human Editing | A great amount of information is processed by editors and is put into related categories. | The amount of information is often growing faster than limited editors can handle. |
| Crawler | The coverage rate is better than can be reached by human editing. | The ranking and relevancy of search results are difficult to decide for such large amounts of information. |
| Popularity | "Hot sites" that attract many people's attention are ranked higher. | Some unpopular sites that could be useful to users would never receive a high ranking. |
| Commercial | Users might find sites that pay more for each click match their search better. | Some web sites that don't pay and are highly related to users' search will never show up on top. |

[12] A crawler search engine mainly has three parts; (1) A spider (also called a "crawler" or a "bot") that follows hypertext links to read every page or representative pages on every web site that wants to be searchable; (2) A program that creates a huge index (sometimes called a "catalog") from the pages that have been read; (3) A program that receives a user's search request, compares it to the entries in the index, and returns results to the user. (whatis.com).

[13] For example, Yahoo gives search results from its own directory first. If it can not find any matching from its directory, it will provide search results from Google, a crawler search engine.

[14] Please see Wasserman and Faust (1994), p.188.

[15] Please see Wasserman and Faust (1994), p.192.

[16] The following matrix contains the data of interconnection among search engines in August 2000. An entry of 1 indicates that the search engine indicated by the row is linked to the one indicated by the column. For example, Alta Vista had a link to number 10 (Look Smart). Ask Jeeves had a link to number 4 (Excite).

| | | 1 | 2 | 3 | 4 | 5 | 6 | 7 | 8 | 9 | 10 | 11 | 12 | 13 | 14 | 15 | 16 | 17 | 18 | 19 |
|---|---|---|---|---|---|---|---|---|---|---|---|---|---|---|---|---|---|---|---|---|
| 1 | AltaVista | 0 | 0 | 0 | 0 | 0 | 0 | 0 | 0 | 0 | 1 | 0 | 0 | 0 | 0 | 0 | 0 | 0 | 0 | 0 |
| 2 | AskJeeves | 1 | 0 | 0 | 1 | 0 | 1 | 0 | 0 | 0 | 0 | 0 | 0 | 0 | 0 | 0 | 0 | 0 | 1 | 0 |
| 3 | DirectHit | 0 | 1 | 0 | 0 | 0 | 0 | 0 | 0 | 0 | 0 | 1 | 1 | 0 | 0 | 0 | 0 | 0 | 0 | 0 |
| 4 | Excite | 0 | 0 | 0 | 0 | 0 | 0 | 0 | 0 | 0 | 0 | 0 | 0 | 0 | 0 | 0 | 0 | 0 | 0 | 0 |
| 5 | Google | 1 | 0 | 0 | 1 | 0 | 0 | 1 | 1 | 0 | 0 | 1 | 0 | 0 | 0 | 0 | 0 | 0 | 0 | 1 |
| 6 | GoTo.com | 0 | 0 | 0 | 0 | 0 | 0 | 0 | 0 | 0 | 0 | 0 | 0 | 0 | 0 | 0 | 0 | 0 | 0 | 0 |
| 7 | HotBot | 0 | 0 | 0 | 0 | 0 | 0 | 0 | 0 | 0 | 0 | 1 | 0 | 0 | 0 | 0 | 0 | 0 | 0 | 0 |
| 8 | Infoseek | 0 | 0 | 0 | 0 | 0 | 0 | 0 | 0 | 0 | 0 | 0 | 0 | 0 | 0 | 0 | 0 | 0 | 0 | 0 |
| 9 | iwon | 0 | 0 | 0 | 0 | 0 | 0 | 0 | 0 | 0 | 0 | 0 | 0 | 0 | 0 | 0 | 0 | 0 | 0 | 0 |
| 10 | LookSmart | 1 | 0 | 0 | 0 | 0 | 0 | 0 | 0 | 0 | 0 | 0 | 0 | 0 | 0 | 0 | 0 | 0 | 0 | 0 |
| 11 | Lycos | 0 | 0 | 0 | 0 | 0 | 0 | 0 | 1 | 0 | 0 | 0 | 0 | 0 | 0 | 0 | 0 | 0 | 0 | 0 |
| 12 | MSN | 0 | 0 | 0 | 0 | 0 | 0 | 0 | 0 | 0 | 0 | 0 | 0 | 0 | 0 | 0 | 0 | 0 | 0 | 0 |
| 13 | Netscape | 0 | 0 | 0 | 0 | 0 | 0 | 0 | 0 | 0 | 0 | 0 | 0 | 0 | 0 | 0 | 0 | 0 | 0 | 0 |
| 14 | NorthernLight | 0 | 0 | 0 | 0 | 0 | 0 | 0 | 0 | 0 | 0 | 0 | 0 | 0 | 0 | 0 | 0 | 0 | 0 | 0 |
| 15 | OpenDirectory | 1 | 0 | 0 | 0 | 1 | 0 | 1 | 1 | 0 | 0 | 0 | 0 | 1 | 1 | 0 | 0 | 0 | 0 | 1 |
| 16 | raging | 0 | 0 | 0 | 0 | 0 | 0 | 0 | 0 | 0 | 0 | 0 | 0 | 0 | 0 | 0 | 0 | 0 | 0 | 0 |
| 17 | Snap | 1 | 0 | 1 | 1 | 0 | 0 | 0 | 0 | 0 | 0 | 0 | 0 | 0 | 0 | 0 | 0 | 0 | 0 | 1 |
| 18 | WebCrawler | 0 | 0 | 0 | 1 | 0 | 0 | 0 | 0 | 0 | 0 | 0 | 0 | 0 | 0 | 0 | 0 | 0 | 0 | 0 |
| 19 | Yahoo | 1 | 0 | 1 | 0 | 1 | 0 | 1 | 1 | 0 | 0 | 0 | 0 | 0 | 1 | 0 | 0 | 0 | 0 | 0 |

[17] The figure was generated with the software package KrackPlot.

[18] We used UCINET, a quantitative software analyzing sociomatrix, to analyze the five sets of matrix data.

[19] 1992 Horizontal Merger Guidelines were jointly issued by The U.S. Department of Justice and Federal Trade Commission.

[20] iWon came into the rating in Oct. 1999 and Raging, came into the rating in May, 2000. The other 17 search engines were on the rating throughout the survey period.

[21] Based on rating of audience reach of search engines from Nielsen/NetRatings in June, 2000. AOL is not included in my sample because it is not listed by Nielsen/NetRatings and it is often considered an ISP company. However, it also provides search service. Therefore, the assumption of that top19 search engines account for 100 % of market shares may be distorted.

[22] The major revenue of search engines comes from advertising sales and is proportional

to their audience reach. The more search engine users visit their sponsor web sites, the more revenue the search engines make. Although advertising sales might be used to calculate market shares more accurately, it is difficult in practice to collect this data because most web sites do not release this information. Therefore, we assume the percentage of audience reach is proportional to the percentage of market sales.

23   Please see note3.

24   The trend of concentration ratio is shown as follows:

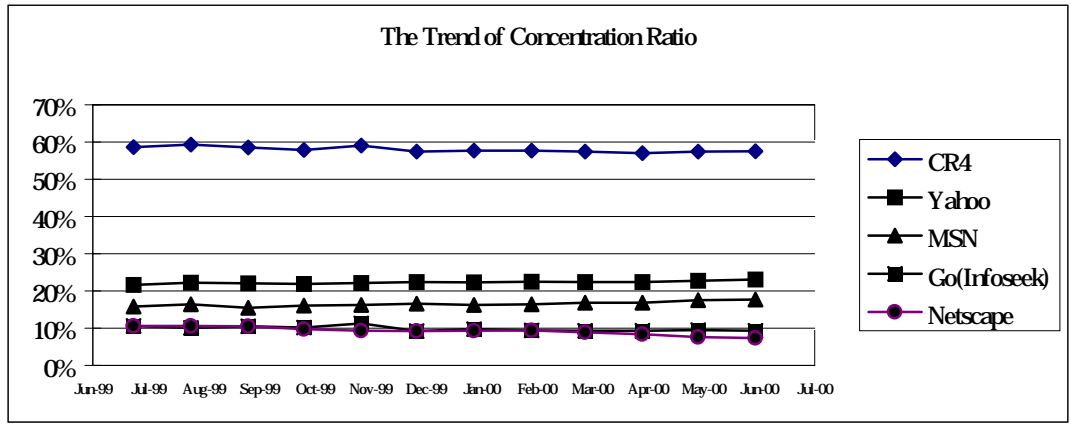

25   The Trend of HHI is shown as follows:

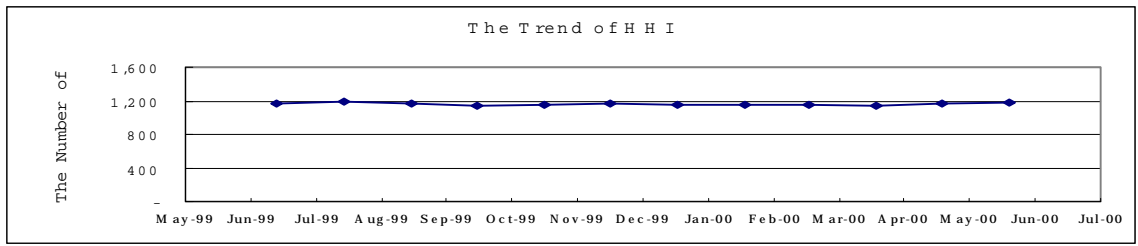

26   The calculation of the NAHHI for August 2000 is listed as follows:

$$\text{NAHHI} = \sum_{i=1}^{N} (100Qi)^2$$

$$\text{Qi(Market Share of Search Engine i)} = \frac{\text{Possible Audience Reach of Search Engine i}}{\sum_{i=1}^{N} \text{Possible Audience Reach of Search Engine i}}$$

Where
Possible Audience Reach of Search Engine i
= Audience Reach of Search Engine i +

$$0.7 \sum_{j}^{J} \text{Audience Reach of the Search Engine j that places a hyperlink pointing to Search Engine i}$$

Assume their audience reach overlap 30%

The indegree data were from Aug. 2000 while the audience reach data were from Jun. 2000 (please see note3). We made an assumption that the audience reach of the search engine i and those search engines that connects to it have 30 % overlap, which means 30 % of their visitors are the same. Based on this assumption, the NAHHI is 870, which indicates an unconcentrated market. In order to reduce the bias of this assumption, we ran a sensitivity analysis varying overlapping rate from 0% to 90%. The NAHHI doesn't change much, still showing an unconcentrated market.

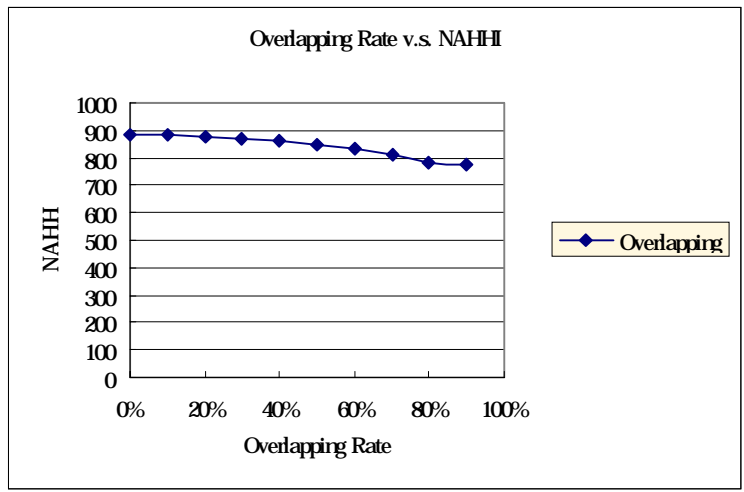

[27] Shapiro and Varian (1998) stated that collective switching cost, the combined switching costs of all users, is a challenge to companies to attract users to a new network. Collective switching costs work in a nonlinear way: convincing ten people in a network to switch to a new network is more than ten times as hard as getting one user to switch. The worst problem is that no one wants to be the first to give up the network externality and risk being stranded.

[28] Microsoft in 1998 spent $400 million buying Hotmail, which was a web site providing free e-mail and had 10 million members as of Jan. 1998. The average cost of getting a member was $40.

[29] Yahoo in 1999 spent $3.5 billion buying Geocities, which was a web site providing free space for publishing web pages and had 3.5 million web sites hosted under it as of Jan. 1999.

[30] In the search engine market, CR4 and HHI were 58% and 1183 as of Jun. 2000. According to 1992 Merger Guidelines, this market is moderately concentrated. When there is a merger producing an increase in the HHI of more than 100 points, post-merger potentially raise significant competitive concerns.

[31] The following table shows the examples of search engines that are bundled with other products or services from the same parent companies.

| Search Engine | Parent Company | Product Bundling |
|---|---|---|
| MSN | Microsoft | The default homepage of IE either links to MSN web site or has a hyperlink pointing to MSN web site. It is estimated that IE dominates 75% of the browser market. Every time when IE is activated, it will show MSN web site unless users change the default web page(more than 60% of users don't change the default homepage.) and MSN site will gain one hit. |

[32] For example, some search engines have presented their sponsors' web sites on the top of search result pages. Users are led to the information that have been selected based on commercial interests.